\documentclass [12pt] {article}
\begin{document}
\begin{center}
\Large{\textbf{Computation of Light Scattering in Young Stellar
Objects}}
\end{center}

\begin{center}
\large{\textbf{P.W.Lucas$^{*}$}}
\end{center}

\small
$^{*}$Department of Physical Sciences, University of Hertfordshire, College Lane,
Hatfield AL10 9AB, UK. email: pwl@star.herts.ac.uk. Fax: +44 1707 284644

\normalsize
\vspace{4mm}

	A Monte Carlo light scattering code incorporating aligned non-spherical
particles is described. The major effects on the flux distribution, linear
polarisation and circular polarisation are presented, with emphasis on the
application to Young Stellar Objects (YSOs). The need for models with non-spherical
particles in order to successfully model polarisation data is reviewed. 
The ability of this type of model to map magnetic field structure in embedded
YSOs is described. The possible application to the question of the origin of 
biomolecular homochirality via UV circular polarisation in star forming 
regions is also briefly discussed.

Author Keywords: Monte Carlo

\section{Introduction}

	The technique of Monte Carlo modelling of light scattering in
spatially resolved nebulae has proven very useful in the field of astrophysics.	
This stochastic method works by generating millions of photons and allowing them 
to propagate in random directions, undergoing single or multiple scattering into 
directions sampled from a theoretical or empirical probability distribution 
function. Those 
photons which eventually escape the bounds of the system are mapped on to a grid
using their location and direction after the last scattering. This grid then 
represents a 2-D image, such as that obtained with a telescope using an optical
CCD camera or an infrared array camera. 

	Monte Carlo modelling is used to investigate the physical structure and
composition of both the dusty nebulae surrounding YSOs, which are stars in the 
process of formation, and protoplanetary nebulae and planetary nebulae (PPNe and 
PNe), which surround stars near the end of their life cycle that have ejected their 
outer atmospheres. Another application is imaging and spectroscopic modelling
of electron scattering for hot stars with an ionised stellar wind
eg. [1],[2] but in this paper only imaging and 
polarimetric imaging models of scattering by dust are considered. 

	Until recently published models considered only spherical dust grains,
using the Mie theory to generate the appropriate amplitude and Stokes matrices
for scattering in different directions. Such models have been successful in
explaining most of the major observed features of YSOs eg. [3],[4],[5],[6],[7],[8].
However several features of the linear and circular polarisation data have not
been successfully reproduced by models using spherical grains (see Section 2), 
so a number of researchers have begun to incorporate scattering by aligned 
non-spherical 
particles (hereafter dichroic scattering) into their codes. Wolf, Voshchinnikov and 
Henning[9] published the first demonstration of such a model, incorporating 
dichroic scattering but not fully treating the effect of dichroic extinction
on the polarisation state. Such a model is strictly applicable only to the optical 
thin case (eg. a planetary nebula).
Whitney and Wolff[10] have made a detailed investigation of the effect
of dichroic scattering and extinction on images and linear and circular polarisation 
maps, applied to spherical clouds and to YSOs for the case that 
the particles are uniformly aligned with a magnetic field along the axis of the 
system. The code presented here is conceptually similar to [10],
so this paper deal mainly with results and details of the algorithm which were
not already covered by those authors.

\section{Limitations of spherical grain treatment and applications of non-spherical grains}

	Models with spherical grains can successfully reproduce most features	
of imaging data but are less successful with polarimetric data. Specific
examples where previous Monte Carlo work either fails to match data or leads to a
misleading match to data are:
	
1) The phase function (distribution of scattered light) in imaging data. Spheres 
and even spheroids tend to overestimate the amount of back-scattering since their
symmetry leads to constructive interference at deflection angles near 180$^{\circ}$
which are unlikely to occur for realistic grains with imperfect symmetry. A 
consequence is that the grain size distribution inferred by fitting an 
observed phase function is likely to include too many large grains (since large
grains back-scatter less). This error may have occurred in recent papers on the dust 
ring around the GG Tau A binary YSO system ([11],[12]).

2) The circular polarisation (CP) produced by spherical grains is very low and arises
only from multiple scattering if the source of radiation is essentially
unpolarised, as is the case for most stars. Observations of high infrared circular
polarisation in YSOs (up to 20\%, eg. [13],[14];[15],[16]) require the presence of 
significantly non-spherical grains, which can produce high polarisation via both 
single and multiple scattering or by dichroic extinction of linearly polarised light.

3) Near infrared linear polarisation (LP). The extended regions of aligned vectors 
seen in a minority of YSOs eg. L1551 IRS5, IRAS 04361+2547, IRAS 04239+2436 and
YLW16a ([8],[17]) have not been reproduced with 
Monte Carlo models using spherical grains. Such models have more success (at least
qualitatively) in reproducing the thin region of aligned vectors or 'polarisation 
disk' seen more frequently in YSOs by a combination of multiple scattering and 
instrumental blurring of the polarisation pattern due to limited spatial resolution.

	In addition, long standing direct evidence for emission from 
non-spherical grains at a range of temperatures and distances from YSOs is 
provided by the many published observations of linear polarisation at wavelengths 
$5< \lambda < 1500~\mu$m (eg. [18],[19]). At these wavelengths scattering becomes 
unimportant due to the small size parameter (and hence low albedo) of grains which 
predominate in the disks and envelopes around YSOs. This has been modelled in detail 
by Aitken et al.[20]. The size parameter is defined as $x=2\pi a/\lambda$.

	The motivation for modelling with non-spherical grains stems from the 
important information that can be extracted 
from polarisation data. In the astrophysical context it is routine to infer the 
direction of the tangential component of the magnetic field in the interstellar 
medium or near a YSO from linear polarisation data. The magnetic field is thought to
play a major role in the star formation process but this is poorly understood. 
Observations of magnetic field structure in the central regions of the circumstellar 
envelope of a YSO could potentially provide valuable tests of different theoretical 
models of star formation via the collapse of a large molecular cloud. While the 
mechanism for 
dust grain alignment is often unclear, eg. [21],[22],[23],[24], it is generally 
accepted that the alignment is such 
that the rotation axis giving the largest moment of inertia is on average 
parallel to the magnetic field. Linear polarimetry in mid-infrared and 
sub-mm wavebands permits a direct measurement of the direction of tangential
component of the magnetic field along the line of sight to a star but it is 
generally not possible to map the magnetic field in circumstellar matter, due to 
poor spatial resolution in the sub-mm and the small size of the emitting region 
in the mid-infrared (see [25] in this issue). In the near infrared YSOs often 
display bright 
extended nebulae which can be observed at high signal to noise. Monte Carlo 
modelling of the linear and circular polarisation produced by dichroic scattering 
and extinction in this waveband is at present the only way that the magnetic field 
structure can be studied on scales of tens to hundreds of astronomical units (AU), 
so this should be pursued despite the complexity of the problem.

	Another important application of this type of Monte Carlo model is to the 
possible connection between high circular polarisation in YSOs and the origin
of biomolecular homochirality [13]. The homochirality of DNA
and amino acids in terrestrial organisms may well have an extraterrestrial 
cause (this is supported by data from meteorites, eg. [26]) 
and it is thought to be a prerequisite for the origin of life. Circularly polarised
light in star forming regions is one possible explanation for an extraterrestrial
origin of homochirality. The relevant waveband
for this hypothesis is the near ultraviolet: photons at these wavelengths can 
preferentially dissociate organic molecules with left handed or right handed (L or D) 
chiral structure, depending on the CP state. It is very difficult to observe YSOs
in this waveband, owing to the very high optical depth of the ambient medium,
so Monte Carlo simulations are the only way to investigate the degree of CP
at the relevant wavelengths.

\section{The code and dust grain model}

	The algorithm incorporates several measures to improve efficiency which have
been developed by others who work with stochastic models, see eg. 
[27], [5] and references therein. In this work we model a homogeneous
sphere and an axisymmetric 2-D disk~$+$~envelope distribution of matter which is 
a simple model of a YSO (see Figure 1). Non-axisymmetric 3-D systems
can be modelled with only a minor modification of the code (see [8]), 
but such models are far more computationally intensive since output photons
must then be binned in the azimuthal ordinate as well as polar viewing angle. A 
unique 3-D solution is hard to identify with 2-D imaging and polarisation data, though
observations at several wavelengths provide some depth perception by penetrating
to a range of optical depths. 

	The 2$\times$2 amplitude matrix used in the phase 
function and the 4$\times$4 Stokes matrix which provides the new Stokes vector of 
each photon after scattering are pre-calculated using the T-matrix codes 
of Mishchenko (see http://www.giss.nasa.gov/$^{\sim}$crmim). The complexity of the 
code is increased compared to the
treatment for spheres because all 4 elements of the amplitude matrix and all
16 elements of the Stokes matrix are non-zero. At present the Monte Carlo code works 
only with oblate spheroids, whose appearance does not change with rotation about the 
axis of greatest inertia. However, it can treat an arbitrary magnetic field structure.
The quadruple precision version of Mishchenko's code 'amplq.new.f' is required for 
ultraviolet work with grains which are highly flattened (at least in the astronomical
context) with axis ratio of 3:1 and have equivalent
surface area sphere radii up to 0.75~$\mu$m, which may be suitable for dust in the 
envelopes around YSOs. For longer wavelengths the double precision version 
'ampld.new.f' is used.
The Stokes matrix elements are then used to calculate the 3 non-zero elements of 
the 4$\times$4 extinction matrix which modifies the polarisation state of each 
photon during flight between scatterings (see [28], [10], [29]). The phase function, 
Stokes matrix
elements and extinction matrix elements are stored as look-up tables in the code,
usually sampled at intervals of 2$^{\circ}$ (essential for ultraviolet work
at large size parameters) or sometimes 4$^{\circ}$ for work at small size 
parameters ($x<1$), appropriate in the infrared.
The Mishchenko code is used to produce matrix elements averaged over a 
distribution of grain sizes for the full range of polar scattering angles 
(deflection angle, $D$) and the full range of azimuthal and polar grain orientation 
angles ($\alpha$, $\beta$). 

The algorithm proceeds as follows:

(i) Millions of 'unpolarised photons' with Stokes vector $(IQUV)=(1,0,0,0)$ are 
generated at random points on the stellar surface or a hot accretion disk and are 
assigned random directions of flight. Real polarised photons could be used
but unpolarised photons are more efficient, representing the average of many
polarised photons, like a classical light wave.

(ii) Each photon steps along its path, gathering optical depth until it is 
absorbed or scattered after a reaching a randomly generated optical depth
($\tau=-ln(1-R)$, for random number $0<R<1$) or else escapes the system. 
To improve efficiency, photons are not allowed the albedo-dependent chance to be 
absorbed and discarded until several scatterings have occurred. Instead the 
'weight' or importance of the photons is reduced by multiplying the Stokes vector by 
the albedo, until Stokes $I<10^{-3}$, at which point the weight is held constant
and absorption occurs is a random manner.

(iii) Treatment of the Stokes vector upon scattering.
The azimuthal scattering angle of the photon, $\gamma$, is set 
at zero for purposes of calculating the new Stokes vector after scattering, since 
$\gamma$ and $\alpha$ are not independent, i.e. the azimuthal grain orientation
is calculated relative to the azimuthal deflection angle. This provides an essential
saving in the size of the Stokes matrix look-up tables, which is permissible since
the Stokes vector is rotated into the scattering plane for the calculation, and then 
rotated back into the systemic frame, as described by Hovenier and van der Mee[30], 
and in [31],[32] :

\vspace{4mm}
\hspace{1cm} $\left(\begin{array}{c}I\\Q\\U\\V \end{array}\right)_{scat} = {\bf L}(\pi-\sigma_2) \hspace{2mm}{\bf Z} \hspace{2mm} {\bf{L}}(-\sigma_1) 
\left(\begin{array}{c}I\\Q\\U\\V \end{array}\right)_{inc} $ \hspace{7cm}(1)

\vspace{4mm}
where the Stokes matrix ${\bf Z} = \left(\begin{array}{cccc}
Z_{11} & Z_{12} & Z_{13} & Z_{14}\\ 
Z_{21} & Z_{22} & Z_{23} & Z_{24}\\
Z_{31} & Z_{32} & Z_{33} & Z_{34}\\
Z_{41} & Z_{42} & Z_{43} & Z_{44}
\end{array}\right)$

\vspace{4mm}
and ${\bf L}$ is the simple rotation matrix described in those papers, acting on the 
rotation angles $\sigma_1$ and $\sigma_2$. These angles and their signs are as 
defined by [30]. The definition of LP position angle is also as defined in that work
and in [29]. The position angle 
is zero along the projection of the symmetry axis of the system on to the plane 
perpendicular to the Poynting vector. Position angle 
increases in a clockwise rotation from the point of view
of the distant observer, so this is a left handed definition if the thumb points 
along the Poynting vector. If the zero angle is regarded as due north this is an 
increase to the west (in the astronomical, upward looking sense of west, opposite 
to the terrestrial sense). This is opposite to the usual astronomical observational 
convention of polarisation increasing from north through east but we retain it for 
consistency with previous modelling work. Positive Stokes V corresponds to left 
handed CP, with the thumb indicating the Poynting vector. This definition follows
that of [29], since the T-matrix codes of Mishchenko are used. This is opposite
to the definition of [30].\footnote{The sense of positive V was erroneously stated as 
right handed in the published version of this paper, but plots of positive and negative 
V remain correct.}
 
(iv) Determination of scattering direction. Upon scattering the polar and azimuthal
deflection angles of the photon ($D$, $\gamma$) are randomly 
sampled from the full scattering phase function by rejection sampling. $\gamma$  
is a left handed rotation about the direction of the Poynting vector, with the
zero of $\gamma$ parallel to the position angle of LP. The new direction of 
flight in the systemic frame is then calculated by transforming to the frame 
defined by the incident photon. The phase function is 
dependent on the grain orientation angles ($\alpha$, $\beta$) and the 
linear and circular polarisation state of the incident photon (see Appendix). 

(v) During flight, the polarisation state of each photon is modified at each 
step by the dichroic extinction of the medium. To do this the Stokes
vector is rotated into the plane defined by the magnetic field (with which the grains are 
aligned) and the direction of flight (Poynting vector), modified as shown below and 
then rotated back into the systemic frame. Thus dichroic extinction by an arbitrary 
magnetic field structure can be treated using only 3 independent non-zero elements of the 
$4\times4$ extinction matrix, after the manner described by [10], 
see eqs.(2-5). For computational purposes this is more efficient than calculating 
the full extinction matrix in the manner expressed by [27], which 
incorporates the rotations. 

\vspace{4mm}
$I_{k+1} = 0.5\times\{(I_k + Q_k).exp[-n (K_{11} + K_{12})  \delta s] + (I_k - Q_k).exp[-n (K_{11} - K_{12}) \delta s]\}$ \hspace{1.15cm} (2)

\vspace{4mm}
$Q_{k+1} = 0.5\times\{(I_k + Q_k).exp[-n (K_{11} + K_{12}) \delta s] - (I_k - Q_k).exp[-n (K_{11} - K_{12}) \delta s]\}$ \hspace{1cm} (3)

\vspace{4mm}
$U_{k+1} = exp[-n K_{11} \delta s].\{U_k cos(n K_{34} \delta s) - V_k sin(n K_{34} \delta s)\}$ \hspace{5.95cm}(4)

\vspace{4mm}
$V_{k+1} = exp[-n K_{11} \delta s].\{V_k cos(n K_{34} \delta s) + U_k sin(n K_{34} \delta s)\}$ \hspace{6cm}(5)

\vspace{4mm}
$K_{11}$, $K_{12}$ and $K_{34}$ are the 3 independent elements of the extinction
matrix, evaluated at grain orientation ($\alpha,\beta) = (0,\beta)$. The $k$ subscript
denotes the $k$th step of the photon along the path, $\delta s$ is the physical 
length of the step and $n$ is the number density of grains.

(vi) Those photons which eventually escape the system are output into file bins
corresponding to the full range of polar viewing angles (0 to 180$^{\circ}$). In 2-D
axisymmetric models all photons are assumed to be output into the observer's
azimuthal viewing angle. This means that the azimuthal coordinate of each 
photon is not specified when the photon is generated and the code only
calculates the {\it change} in azimuthal coordinate during flight.
For systems which have mirror symmetry about the equatorial plane the equivalence 
of the output into the northern and southern hemispheres is a useful test of the code.
A separate program is then used to map the photons on to a 2-D image grid, using the
location and direction of the photon at last scattering. For each grid element the
the 4 Stokes components are summed for all photons mapped on to that element. This
generates the raw $(I,Q,U,V)$ model output maps. For comparison with real data the raw
$(I,Q,U,V)$ maps are then convolved with a point spread function appropriate for the
observations. Accurate measurement and selection of the point spread function
is crucial for useful modelling in cases where the optical depth to the central
star $\tau \le 6$, since the strong central flux peak (which is often highly 
polarised by dichroic extinction) tends to modify or obscure features arising from 
scattered flux.

	This algorithm is very similar to that developed in [10] and both groups
chose to employ the Mishchenko code to provide the input matrices. 
The codes of the two groups have been bench tested against each other for the case of 
the uniform axial magnetic field, applied to both uniform spherical clouds and 
axisymmetric 2-D disk~$+$~envelope distributions of matter suited to YSOs.
The only significant known differences are: (a) the algorithm in [10] can handle 
precessing oblate or prolate spheroids, whereas this code uses only perfectly 
aligned oblate 
spheroids at present; (b) this algorithm can handle arbitrary magnetic field
structures, whereas [10] dealt only with the case of an axial field.
Any other differences
are believed to be trivial, though the source codes have not been compared. The 
precision of the code is determined by the length of photon steps in stage (ii) and the 
sampling frequency of the look up tables. It can be altered to suit the desired 
spatial resolution and angular binning of the output maps. The necessary precision
for a given application is best determined empirically. 
We note that bench testing is highly desirable for computations with non-spherical
particles, since the many of the symmetries in photon output which can be relied 
upon to test computation with spheres no longer apply, particularly if the magnetic
field is non uniform.
The algorithm was coded in Fortran 77, and can be run in series or in parallel.
Nearly 100\% parallelisation efficiency is achieved by evenly distributing the
photons between CPUs. The parallel code is compiled with the Silicon Graphics MIPSpro 
Fortran 77 compiler but it should run in series with any compiler compatible with
Fortran 77. The program speed depends mainly on the complexity of the structure 
of circumstellar matter: for complicated systems like a YSO the move from spherical 
grains to non-spherical grains does not significantly increase the run time.
The large look up tables lead to far greater RAM requirements, however, and a sampling
of the matrices at intervals of 1$^{\circ}$ would be beyond the capacity of most
desktop PCs at present.

All simulations shown in this paper are run for a wavelength of 
2.2~$\mu$m and a grain mixture consisting of (1) silicate particles
(n=1.7 - 0.03i) with equivalent surface area radii in the range 0.005 to 
0.35~$\mu$m, and (2) small amorphous carbon particles[33] (n=2.77-0.80i) in the size 
range 0.005 to 0.030~$\mu$m. All particles are oblate with a 3:1 axis ratio. 
For both grain types the size distribution is given by $n(r) \propto r^{-3.5}$.
The maximum grain size of 0.35~$\mu$m is towards the lower end of the plausible 
range in YSOs, comparing with the widely adopted maximum size of 0.25~$\mu$m for 
interstellar dust. The choice of a mixture of grains with fairly small size 
parameters contributes to the exotic behaviour of CP in the results shown here 
(see Section 4.2).
The scattering is dominated by the shiny silicates but the absorptive 
amorphous carbon particles serve to reduce the albedo to approximately 0.16.
This mixture is used as a compromise between possible dielectric and highly 
absorptive mixtures.
The true composition of circumstellar dust near YSOs is uncertain but is known
to include water ice and CO ice (eg. [34]), believed 
to be deposited in mantles around a silicate core. The ices are likely to have 
only a small imaginary component of refractive index, like the silicate used here,
but the mantles may have metallic inclusions. In the models shown here the opacity 
of the grain mixture is set to unity and absorbed into the product of opacity
and the mass density of the medium.

\section{Model Results}

\subsection{Homogeneous Sphere with a uniform magnetic field}

	The spatially resolved (I,Q,U,V) map of a 
sphere illuminated by a point source at its centre has 
been discussed by [10] for the case
of 2:1 prolate and oblate spheroids with varying degrees
of alignment. Their results are reproduced by this code
(considering only perfectly aligned oblate grains, which
show the strongest effects of non-spherical symmetry).
Here some additional features arising from 3:1 grains are
noted which arise in the case of fairly high optical depth.

{\bf Stokes I: A sphere can appear as a bipolar nebula.} 
In [10] it is described how an optically
thin sphere appears slightly elongated along the axis of the 
magnetic field, due to the greater optical depth and hence larger scattered
flux along the axis (Figure2(a)). For the case of high optical 
depth the effect of enhanced scattering by grains located along 
the axis is outweighted by the strong extinction in that direction,
which prevents light from reaching the upper and lower
parts of the sphere, as shown in Figure 2(b). Thus the sphere
appears fairly round near an optical depth of unity and the 
image becomes elongated perpendicular to the magnetic field at high 
optical depth. Here it is shown that for 3:1 grains with
an optical depth $\tau=10$ along the axis from the source to 
the edge of the sphere, the greater extinction along the
axis squashes the flux distribution so much that it becomes
bipolar. In this model the optical depth perpendicular to the
axis is $\tau=2.9$ from source to edge of sphere.
The bipolar structure begins to appear at $\tau=5$.

{\bf Stokes Q and U:} For flattened grains the linear 
polarisation pattern departs from the centrosymmetric
pattern commonly observed in circumstellar nebulae. In
this model (Figure 3(a-b)) vectors oriented along the magnetic field
axis are defined as having a position angle of zero, i.e. positive
Q and zero U. The scattered electric vector is enhanced in the 
plane containing the two long axes of the grains, leading to
enhanced negative Q at low optical depth (i.e. vectors perpendicular 
to the axis). At high optical
depth the effect of dichroic extinction becomes more important, 
preferentially absorbing photons with a negative Q, and leading 
to a vector pattern dominated by positive Q. The Stokes U 
component is absorbed more than positive Q, so at high optical
depth the vectors have an axial orientation, parallel to the 
magnetic field. This is described in more detail by [10].

{\bf Stokes V and U: The sense of circular polarisation can 
reverse several times between the centre and edge of the sphere}
Non-spherical grains readily produce high levels of CP. In the
more familiar case of spherical grains the CP is weaker and the 
pattern is quadrupolar: the sense of CP alternates between adjacent
quadrants of the image of a spherical cloud, with zero CP along 
the vertical and horizontal diameters. This has been observed
in several YSO nebulae (eg. [35]) and is simply
due to alternations in the sign of the Stokes U component as 
perceived in the scattering plane.

At low optical depth, spheroidal grains produce the same quadrupolar
pattern as spheres (Figure 4(a)). However at high optical depth 
dichroic extinction plays a prominent role in modifying the Stokes
vector after scattering (equns. (4-5)). 
In the case shown (Figure 4(b-c) the CP produced by scattering is quite high 
so V/I is large at the edges of the image, where dichroic extinction has
little effect on the polarisation state due to the short path length to the
edge of the system after scattering. The CP falls as the line of sight moves 
inward due to conversion of Stokes V to Stokes U as the path length 
for dichroic extinction of the scattered light increases. If the
path length is long enough (as in Figure 4(b-c)) then Stokes V will 
reverse in sign one or more times between the centre
and edge of the image as Stokes V is converted to U, then to negative
V, then to negative U, then back to positive V and so on.
The behaviour of Stokes U is easy to understand. Dichroic extinction 
introduces a phase difference between the electric field components
parallel and perpendicular to the magnetic field axis, generating CP.
CP is maximised when the phase difference, $\delta$, is 90$^{\circ}$, 
declines to zero when $\delta= 180^{\circ}$, and has the opposite
sign for $\delta > 180^{\circ}$, peaking at $270^{\circ}$ and so
on at $450^{\circ}$ and $630^{\circ}$ etc. Reversals in Stokes V were 
also noted by [10] for the case of 2:1 grains, though their 
origin was not discussed. The effect of the reversals in U on the LP 
vector plot (Figure 3(b)) is fairly subtle: the vector pattern is 
dominated by Q at high optical depth, so the vectors merely twitch 
slightly to either side of the vertical as the sign of U changes.

While Figure 4(b-c) shows the effects of mutual interconversion of 
Stokes U and Stokes V initially produced by scattering, there may be 
cases where scattering produces little CP, eg. if grains are not aligned, 
or if the Z$_{41}$ Stokes matrix element happens to be small. The observed CP 
may then be produced almost entirely by dichroic extinction by aligned grains
in the foreground.

In summary, CP can be produced either by scattering of initially
unpolarised light (both single and multiple scattering) or by 
dichroic extinction of linearly polarised light. At high optical 
depth the latter is likely to be the more important mechanism.

\subsection{A Young Stellar Object with a uniform axial magnetic field}

	Model results for an embedded YSO are shown in Figure 5(a-d).
The circumstellar matter has the following components, illustrated in Figure 1.

(i) an accretion disk of radius 400~AU in vertical hydrostatic equilibrium, with 
density as in [3] (their eq.(5)), for the case of steady accretion. For such a
disk the volume density declines with radius as $r^{-15/8}$ and and the scale 
height increases with radius $r^{9/8}$. The disk also has a small inner hole of 
radius 2$\times10^{10}$~m, which is a factor 20 larger than the adopted stellar 
radius.

(ii) a large, equatorially condensed envelope of radius 5000~AU. This has the
following density distribution:

\vspace{4mm}
\hspace{1cm}$\rho_{env} = C \hspace{1mm} R^{-3/2} (1/(\mu^k + 0.05))$ \hspace{1cm}

\vspace{4mm}
where C is a free parameter (3.7$\times 10^6$~Kg m$^{-3/2}$ in this case),
$R$ is the 3-D radius for the natural system of cylindrical polar coordinates 
defined by the disk and rotation axis ($R=\sqrt{r^2+z^2}$) and $\mu = |z/R|$.
The index $k$ determines the degree of equatorial condensation and is set to 2.0
here. This density distribution was derived empirically following previous
comparison of observations and models[7,8]. Many other algebraic forms could be 
chosen.

(iii) an evacuated conical bipolar cavity with opening angle of 90$^{\circ}$
and radius of 50~AU in the disk plane.

	The behaviour of the four Stokes components is similar to that seen in 
the homogeneous sphere at high optical depth. The main
differences are: (a) that the image is less strongly affected
by the non-spherical shape of the grains, since its bipolar or cometary
appearance [4,6,8] is determined more by the envelope and cavity 
structure of the system (Figure 1); (b) the radial sign changes in Stokes U 
and V between the centre and edge of the image are confined to a small 
central region, since the density of the envelope 
(and hence optical depth) is greatest near the centre of the system. 
The azimuthal sign changes between quadrants remain prominent however.

	The radial sign reversals in Stokes U occur at slightly larger radii than 
those of Stokes V, but would be difficult to observe since the LP pattern
is dominated by Stokes Q in the central region. The radial sign reversals in
Stokes V are certainly observable in principle via imaging
polarimetry at high spatial resolution, eg. with Adaptive Optics.
However, the sign reversals will only be seen if the grains have a fairly large 
axis ratio, so that the dichroic component of optical depth ($\int n(s) 
K_{34} ds$) reaches $\pi/2$ before the total optical depth 
($\int n(s) K_{11} ds$) becomes high enough to make observation 
impossible. This is is most likely to occur for dielectric or weakly absorptive
substances at size parameters $x<1$, where the ratio $K_{34}/K_{11}$ is high, and 
may exceed unity.
The number of radial sign reversals in U and V is reduced if the maximum size of 
the silicate grains is increased from 0.35 to 0.75~$\mu$m (not shown) but the 
reversals still occur.
The grains would also need to be well aligned with a magnetic field which is
reasonably uniform on the observed spatial scales. Dichroic
extinction appears to be ubiquitous in YSOs, judging by observation
of the less embedded systems such as T Tauri stars where the central 
source is seen directly (eg. [36]). The degree of LP
produced by dichroic extinction is usually $<10\%$ in T Tauri stars,
but can be much higher in more embedded systems, exceeding 30\% 
in extreme cases such as IRAS 04361+2547 (also known as TMR-1),
see [7]. In such systems it is possible that 
these sign reversals could be observed, but if not their absence
will act as a constraint on models of the dust grain shapes
and the density of the nebula.

\subsection{Changing the magnetic field: a YSO with a uniform toroidal field}

	The structure of the magnetic field in YSO nebulae is expected to be 
axial on scales of thousands of AU due to preferential collapse
of the parent molecular cloud core along the field lines towards
a flattened envelope and disk structure. This is confirmed by 
observations of LP produced by emission from non-spherical grains
at millimetre and sub-mm wavelengths[18,19], as noted in Section 2.
At smaller distances from the protostar the magnetic field is likely to be 
pinched in by the infall of the circumstellar envelope and may also
be twisted by the rapid rotation of the inner accretion disk, eg. [37,38].

	Here we simply show (Figure 6) that for a uniform toroidal magnetic field
structure the LP vector pattern near the centre is rotated through
$90^{\circ}$ relative to the axial magnetic field case so that the vectors 
mostly lie in the plane of the accretion disk. This is the expected result
for LP dominated by dichroic extinction. In reality the magnetic field
may only become toroidal very close to the centre of the system. Test
models (not shown) indicate that in such a case this inner magnetic field
structure can lead to a rotation of the LP vectors in the middle of
the image, relative to those further out, providing a useful diagnostic
of magnetic field structure. However, in the case of high optical depth (i.e. the 
younger, more embedded YSOs) this inner structure might be obscured from 
view by the dichroic extinction of grains aligned with the nearly axial 
magnetic field in the outer parts of the envelope. This topic will be 
investigated in future papers by fitting models to data on 
individual YSOs.

\subsection{Implications for the connection with biomolecular homochirality}

	The Stokes V plot for a YSO Figure 5(c) shows that high degrees
of CP can be emitted in some directions, as required for efficient
asymmetric photolysis of prebiotic molecules in star forming regions.
However, Figure 5(c) is an infrared simulation. At UV wavelengths the
phase function for scattering by sub-micron sized grains becomes highly 
forward throwing, and the flux emerging from the axisymmetric model
system is dominated by photons which have been deflected through small
angles and therefore have weak CP and LP (only a few percent after
single scattering). The axisymmetry of the idealised 2-D system will
also produce zero net CP due to the opposite contributions from adjacent
quadrants but this is a not a serious problem since many asymmetric nebulae
are known eg. Cha IRN [39][40].
The hypothesis of Bailey[13] seems to require fairly high (though at 
present poorly quantified) degrees of UV CP. Hence, a contribution to CP from the 
dichroic extinction mechanism discussed here may be required, which 
will be a topic for a future investigation. Another possibility is generation
of CP by optically active carbonaceous molecules, perhaps associated with the 
almost ubiquitous $2175 \AA$ interstellar absorption peak.

\section{Conclusions}

	Monte Carlo models incorporating dichroic scattering and extinction 
are likely to prove a valuable tool for the study of dusty nebulae. The results 
shown here are in good agreement with recently published results using a similar
algorithm [10], though the larger axis ratio employed here (3:1 rather than 2:1) 
leads to some interesting new results at high optical depth, caused by dichroic 
extinction. The large
number of model parameters (eg. grain size distribution, shape, degree of 
alignment and any spatial variation in these) means that it will often be 
necessary to make run a large number of simulations in order to fit 
observational data. The model polarisation maps shown here contain many
complicated features, even though very simple magnetic field structures 
have been used. Further abstract investigation of the effects of non-spherical
grains therefore seems warranted, as well as comparison with observation.

\vspace{4mm}
{\large Acknowledgements}

\vspace{4mm}
	I wish to thank Tim Gledhill, Allan McCall and Antonio Chrysostomou
for helpful discussions about dichroic scattering. I also thank Barbara Whitney
and Mike Wolff for invaluable bench testing and discussion of the Monte Carlo 
codes. The simulations shown here were run on MIRACLE, the SGI Origin 2000 computer 
at the Hiperspace Centre at University College, London. The author acknowledges
support by the UK Particle Physics and Astronomy Research Council (PPARC).

\section{Appendix - the Phase Function for Spheroids}

	The phase function for spheroids may be derived from
the first line of the Stokes matrix:

\vspace{4mm}
(A1) $I_{sc} = Z_{11}I + Z_{12}Q + Z_{13}U + Z_{14}V$
\vspace{4mm}

where the (real) Stokes matrix elements $Z_{ij}(\alpha,\beta,D,0)$ are functions of
grain azimuthal orientation angle $\alpha$ (defined as a right-handed rotation 
about the Poynting vector relative to
the azimuthal deflection angle, $\gamma$), the grain polar orientation 
angle $\beta$ and photon deflection angle $D$.
The Stokes matrix elements $Z_{ij}$ are formed from non-linear
combinations of the complex amplitude matrix elements $S_{ij}(\alpha,\beta,D,\gamma)$,
with $\gamma$ set to zero in the $S_{ij}$ and $Z_{ij}$ elements.
(We adopt the sign conventions of [29] for the matrix elements since these are used 
in the Mishchenko code.):

\vspace{4mm}$Z_{11} = 0.5 \times (|S_{11}|^2 + |S_{12}|^2 + |S_{21}|^2 + |S_{22}|^2)$;
\hspace{4mm}$Z_{12} = 0.5 \times (|S_{11}|^2 - |S_{12}|^2 + |S_{21}|^2 - |S_{22}|^2)$

\vspace{4mm}
$Z_{13} = -Re(S_{11}S_{12}^{*} + S_{22}S_{21}^{*})$;\hspace{32mm}
$Z_{14} = -Im(S_{11}S_{12}^{*} - S_{22}S_{21}^{*})$

\vspace{4mm}
The coordinate system of [30] is used so $\gamma$ is a left handed 
rotation about the Poynting vector. Since $\gamma$ is set to zero in the matrix 
elements, the orientation of the Stokes vector must be defined relative to $\gamma$,
with $\gamma=0$ corresponding to deflection in the plane containing the linear 
polarisation vector and Poynting vector of the incident photon. 
Again, following [30], the Stokes vector is a left handed rotation about the 
Poynting vector.

This leads to the relations: 

\vspace{4mm}
$Q=I \times LPcos(2\gamma)$   \hspace{1cm} $(LP= \sqrt{Q^2 + U^2}/I)$

$U=-I \times LPsin(2\gamma)$

$V=I \times CP$		\hspace{2.4cm}       $(CP= V/I)$  	

\vspace{4mm}
These relations between the three azimuthal angles ($\alpha$, $\gamma$ and direction
of LP) in the reference frame of the incident photon are illustrated in Figure 7. 

Eq.(A1) then rearranges to:

\begin{eqnarray*}\textmd{(A2)} \hspace{2mm}I_{sc}d\Omega & = & I_{in}\{(1-LP)/2 \times 
(|S_{11}|^2 + |S_{12}|^2 + |S_{21}|^2 + |S_{22}|^2)\\
 &   & +\hspace{2mm}LP\times[(|S_{22}|^2 + |S_{12}|^2)sin^2(\gamma) + (|S_{11}|^2 + |S_{21}|^2)cos^2(\gamma)\\ 
 &   & \hspace{1.2cm}+ sin(2\gamma)Re(S_{11}S_{12}^{*} + S_{22}S_{21}^{*})]\\
 &   & -\hspace{2mm}CP\times Im(S_{11}S_{12}^{*} - S_{22}S_{21}^{*})\}d\Omega\end{eqnarray*}

giving the scattered intensity per unit solid angle $d\Omega$.

Eq.(A2) may also be derived directly from the amplitude matrix by
separating the electric field amplitudes into components parallel
and perpendicular to the scattering plane, after the method
used by [5] to derive the simpler phase function
appropriate for Mie scattering by spheres.

Eq.(A2) is equivalent to the following probability density function:

\vspace{4mm}
\begin{eqnarray*}\textmd{(A3)} \hspace{2mm}P(D,\gamma)dD d\gamma &=& sin(D)\{(1-LP)/2 \times (|S_{11}|^2 + |S_{12}|^2 + |S_{21}|^2 + |S_{22}|^2) \\
 &   & +\hspace{2mm}LP\times[(|S_{22}|^2 + |S_{12}|^2)sin^2(\gamma) + (|S_{11}|^2 + |S_{21}|^2)cos^2(\gamma)\\
 &   & \hspace{1.2cm}+ sin(2\gamma)Re(S_{11}S_{12}^{*} + S_{22}S_{21}^{*})]\\
 &   & -\hspace{2mm}CP\times Im(S_{11}S_{12}^{*} - S_{22}S_{21}^{*})\}dD d\gamma/I_{max}(\alpha,\beta,LP,CP)\end{eqnarray*} 

where $I_{max}(\alpha,\beta,LP,CP)$ is the intensity at the most probable scattering
direction ($D,\gamma$) which is different for each combination of $(\alpha,\beta,LP,CP)$.
This term is used to normalise the probability density function relative to unity.

Eq.(A3) is sampled by rejection sampling: values of both $D$ and $\gamma$
are randomly generated and then accepted if $P(D,\gamma)$ is greater than a 
third random number R ($0<R<1$). This method requires calculation of 
$I_{max}(\alpha,\beta,LP,CP)$ for each possible combination
of incident ($\alpha$, $\beta$, LP and CP). Eq.(A2) must often therefore be 
calculated for all values of all six variables to do this, and the
peak values stored in a look-up table. For large size parameters it may be
assumed that the most probable deflection angle will always be small,
which reduces the number of calculations required. Rejection sampling
is inefficient for highly forward throwing phase functions. However, even at size
parameters up to 20 the run time of the code is not affected, since
it is dominated by the computation of the photon location and Stokes
vector as it moves in many small steps along its path, not by the 
scattering calculation.

Eq.(A3) is usually sampled every every 2$^{\circ}$ in each of the 4 angles, every
2\% in LP and every 20\% in CP, which generally has only a small
effect on the phase function via conversion of some of the
incident Stokes V to the direction-dependent Stokes U. This leads
to 150 billion calculations of Eq.(A2), which takes about an hour on an
SGI Origin 2000 computer, running with several CPUs in parallel.
This speed is achieved by simplifying Eq.(A2) by pre-calculating components
of the equation at the same time that the matrix elements are first
calculated using a slightly modified version of the Mishchenko code, 
ampld.new.f. For small size parameters ($x<1$) the 4 angles and LP may be 
sampled a factor of 2 less often, which reduces the number of iterations of 
A2 by a factor of 32.

\section{References}

[1] Woods, JA. Winds in Cataclysmic Variables. D. Phil thesis, Univ. Oxford, 1991.

[2] Knigge C, Woods JA, Drew JE. The application of Monte Carlo methods to the synthesis of spectral line profiles arising from accretion disc winds. Monthly Not Royal Astron Soc 1995, 273, 225

[3] Whitney BA, Hartmann L. Model scattering envelopes of young stellar objects. I - Method and application to circumstellar disksAstrophys J 1992, 395, 529

[4] Whitney BA, Hartmann L. Model scattering envelopes of young stellar objects. II - Infalling envelopes. Astrophys J 1993, 402, 605

[5] Fischer O, Henning Th, Yorke HW. Simulation of polarization maps. 1: Protostellar envelopes. Astron Astrophys 1994, 284, 187

[6] Fischer O, Henning Th, Yorke HW. Simulation of polarization maps. II. The circumstellar environment of pre-main sequence objects. Astron Astrophys 1996, 308, 863

[7] Lucas PW, Roche PF. Butterfly star in Taurus: structures of young stellar objects.
Monthly Not Royal Astron Soc 1997, 286, 895

[8] Lucas PW, Roche PF. Imaging polarimetry of class I young stellar objects. 
Monthly Not Royal Astron Soc 1998, 299, 699

[9] Wolf S, Voshchinnikov NV, Henning Th. Multiple scattering of polarized radiation 
by non-spherical grains: First results. Astron Astrophys 2002, 385, 365

[10] Whitney BA, \& Wolff M. Scattering and Absorption by Aligned Grains in 
Circumstellar Environments. Astrophys J 2002, 574, 205

[11] Silber J, Gledhill TM, Duchêne G, Ménard F. Near-Infrared Imaging Polarimetry 
of the GG Tauri Circumbinary Ring. Astrophys J 2000, 536, L89

[12] Krist JE, Stapelfeldt KR, Watson AM. Hubble Space Telescope/WFPC2 Images of the 
GG Tauri Circumbinary Disk. Astrophys J 2002, 570, 785

[13] Bailey J, Chrysostomou A, Hough JH, Gledhill TM, McCall A,
Clark S, Menard F, Tamura M. Circular Polarization in Star-Formation Regions: 
Implications for Biomolecular Homochirality. Science 1998, 281,672

[14] Chrysostomou A, Gledhill TM, Menard F, Hough JH, Tamura M, 
Bailey J. Polarimetry of young stellar objects - III. Circular polarimetry of OMC-1. 
Monthly Not Royal Astron Soc 2000, 312,103

[15] Menard F, Chrysostomou A, Gledhill TM, Hough JH, Bailey J. High Circular Polarization in the Star Forming Region NGC 6334: Implications. In: Lemarchand G, Meech K, editors. Bioastronomy 99: A New Era in the Search for Life in the Universe. proc. ASP Conf. Proc. 2000, vol 213, p355. (San Francisco: ASP)

[16] Clark S, McCall A, Chrysostomou A, Gledhill T, Yates J, Hough J. Polarization models of young stellar objects - II. Linear and circular polarimetry of R Coronae 
Australis. Monthly Not Royal Astron Soc 2000, 319, 337

[17] Aspin C, Casali MM, Walther DM. Infrared Imaging and Polarimetry of Star Forming Regions. In: Reipurth B, editor. Low Mass Star Formation and Pre-Main Sequence Objects. ESO Conf. Work. Proc. 1989, 33, p349

[18] Aitken DK, Wright CM, Smith CH, Roche PF. Studies in mid-infrared spectropolarimetry. I - Magnetic fields, discs and flows in star formation regionsMonthly Not Royal Astron Soc 1993, 
262, 456

[19] Tamura M, Hough JH, Hayashi SS. 1 Millimeter Polarimetry of Young Stellar 
Objects: Low-Mass Protostars and T Tauri Stars. Astrophys J 1995, 448, 346

[20] Aitken DK, Efstathiou A, McCall A, Hough JH. Magnetic fields in discs: what can be learned from infrared and mm polarimetry?. Monthly Not Royal Astron Soc 2002, 329, 647

[21] Purcell EM. Suprathermal rotation of interstellar grains. Astrophys J 1979, 231, 404

[22] Roberge WG. Grain Alignment in Molecular Clouds. In: Roberge WG, Whittet DCB, editors. Polarimetry of the Interstellar Medium. ASP conf. series 1995, vol.97, p401.

[23] Lazarian A. Mechanical Alignment of Suprathermal Grains. In: Roberge WG, 
Whittet DCB, editors. Polarimetry of the Interstellar Medium. ASP conf. series 1995, 
vol.97, p425.

[24] Lazarian A. Physics of Grain Alignment. In: Franco J, Terlevich L, López-Cruz O, Aretxaga I, editors. Cosmic Evolution and Galaxy Formation: Structure, Interactions, 
and Feedback. ASP Conf. Proc. 2000, vol. 215, p69.

[25] Hough JH., Aitken DK. Polarimetry in the infrared: what can be learned?. 
J Quant Spectrsoc Radiat Transfer, this issue, 2002

[26] Cronin JR, Pizzarello S. Enantiomeric excesses in meteoritic amino acids. 
Science 1997, 275, 951

[27] Whitney BA. Thomson scattering in a magnetic field. PhD thesis, 
Wisconsin Univ., Madison, 1989.

[28] Martin PG. Interstellar polarization from a medium with changing grain alignment.
Astrophys J 1974, 187, 461

[29] Mishchenko MI, Hovenier JW, Travis LD. 
Light Scattering by Nonspherical Particles: Theory, Measurements, and 
Applications.  Academic Press, San Diego, 2000.

[30] Hovenier JW, van der Mee CVM. Fundamental relationships relevant to the 
transfer of polarized light in a scattering atmosphere. Astron Astrophys 1983, 128, 1

[31] Voshchinnikov NV, Karjukin VV. Multiple scattering of polarized radiation in 
circumstellar dust shells. Astron Astrophys 1994, 288, 883

[32] Code AD, Whitney BA. Polarization from scattering in blobs. Astrophys J 1995, 441, 400

[33] Preibisch Th, Ossenkopf V, Yorke HW, Henning Th. The influence of ice-coated grains on protostellar spectra. Astron Astrophys 1993, 279, 577

[34] Pollack JB, Hollenbach D, Beckwith S, Simonelli DP, Roush T, Fong W. Composition and radiative properties of grains in molecular clouds and accretion disks.
Astrophys J 1994, 421, 615

[35] Chrysostomou A, Menard F, Gledhill TM, Clark S, Hough JH,
McCall A, Tamura M. Polarimetry of young stellar objects - II. Circular polarization 
of GSS 30. Monthly Not Royal Astron Soc 1997, 285, 750

[36] Sato S, Tamura M, Nagata T, Kaifu N, Hough J, McLean IS, Garden RP, 
Gatley I. Infrared polarimetry of dark clouds. II - Magnetic field structure in the 
Rho Ophiuchi dark cloud. Monthly Not Royal Astron Soc 1988, 230, 321

[37] Galli D, Shu FH. Collapse of Magnetized Molecular Cloud Cores. I. Semianalytical Solution. Astrophys J 1993, 417,220

[38] Galli D, Shu FH. Collapse of Magnetized Molecular Cloud Cores. II. Numerical Results. Astrophys J 1993, 417,243

[39] Ageorges N, Fischer O, Stecklum B, Eckart A, Henning Th. The Chamaeleon Infrared Nebula: A Polarization Study with High Angular Resolution. Astrophys J 1996, 463, L101

[40] Gledhill TM, Chrysostomou A, Hough JH. Linear and circular imaging polarimetry of the Chamaeleon infrared nebula. Monthly Not Royal Astron Soc 1996, 282, 1418

\section{Figure Captions}

Figure 1: Schematic diagram of a Young Stellar Object. The magnetic field is
vertical and parallel to the rotation axis of the system. The density
of the envelope declines monotonically with distance from the protostar
and distance from the disk plane. The density spans several orders of 
magnitude, as indicated by the wrapped greyscale shading. The cavity is 
taken to be evacuated.\\

Figure 2: Images of a Homogeneous Sphere with vertical magnetic field. 
(a) $\tau=0.5$, showing slight vertical elongation. (b) $\tau=10$,
showing horizontal elongation and a bipolar structure in the central
region. Optical depth is less along horizontal paths than vertical paths due 
to 3:1 oblate shape of the aligned grains.\\

Figure 3: Linear Polarisation vector maps for a Homogeneous Sphere.
(a) $\tau=0.5$, showing LP dominated by dichroic scattering. (b) $\tau=5$, 
showing LP dominated by dichroic extinction. The magnetic field is vertical and the
long axes of the grains are horizontal. 100\% polarisation corresponds to a 
vector length of 0.8 pixels. The direct flux from the central source at the 
(0,0) position is not included here.\\

Figure 4: Stokes V and U reversals for a Homogeneous Sphere. Bright regions
correspond to positive U and V, dark regions to negative U and V.
(a) at $\tau=0.5$ a simple quadrupolar pattern is seen. This is
similar to that produced by spheres but the CP is much higher,
reaching 30\%. (b) A denser sphere ($\tau=3$) produces a radial reversal
in the sense of CP due to dichroic extinction. CP oscillates between
0 and 20\%, near the middle, and exceeds 40\% at the edge. 
(c) Stokes U for the ($\tau=3$) case, illustrating the exchange between 
U and V, giving 2 radial reversals in the sign of U.\\

Figure 5: A YSO with a uniform axial magnetic field (see Figure 1). The axis ratio of 
the grains remains 3:1. The YSO is 
viewed in the equatorial plane. and appears bipolar in Stokes I due to the high 
optical depth of the envelope in that plane. The image (a) shows geometrical
limb brightening at the walls of the evacuated cavity. The vector map (b)
shows vertical vectors near the middle due to strong dichroic extinction 
of light in the dense central regions. The pattern is more centrosymmetric near
the edges, being dominated by scattering there. (c) and (d) Stokes V and U 
both reverse sign over short distances in the dense central region. The effect of the 
reversals in U can be seen as the radial vectors in (b) approximately halfway
from the middle to the edge of the plot. Raw, unsmoothed model output is 
shown here to reveal the fine spatial details. CP exceeds 50\% at peak.\\

Figure 6: A YSO with a uniform toroidal (horizontal) magnetic field. The vector 
pattern in the central region is rotated through $90^{\circ}$ relative to the 
axial field case (Figure 5(b)).\\ 

Figure 7: Illustration of some of the angles associated with the phase function.
The polarisation vector, azimuthal grain angle and azimuthal deflection angle
lie in the plane perpendicular to the incident photon's Poynting vector. $\alpha$
is a right handed rotation but $\gamma$ is a left handed rotation.

\pagebreak
\section{Figures}

\vspace{2cm}

\begin{figure*}[thbp]
\begin{center}
\begin{picture}(200,210)

\put(0,0){\includegraphics{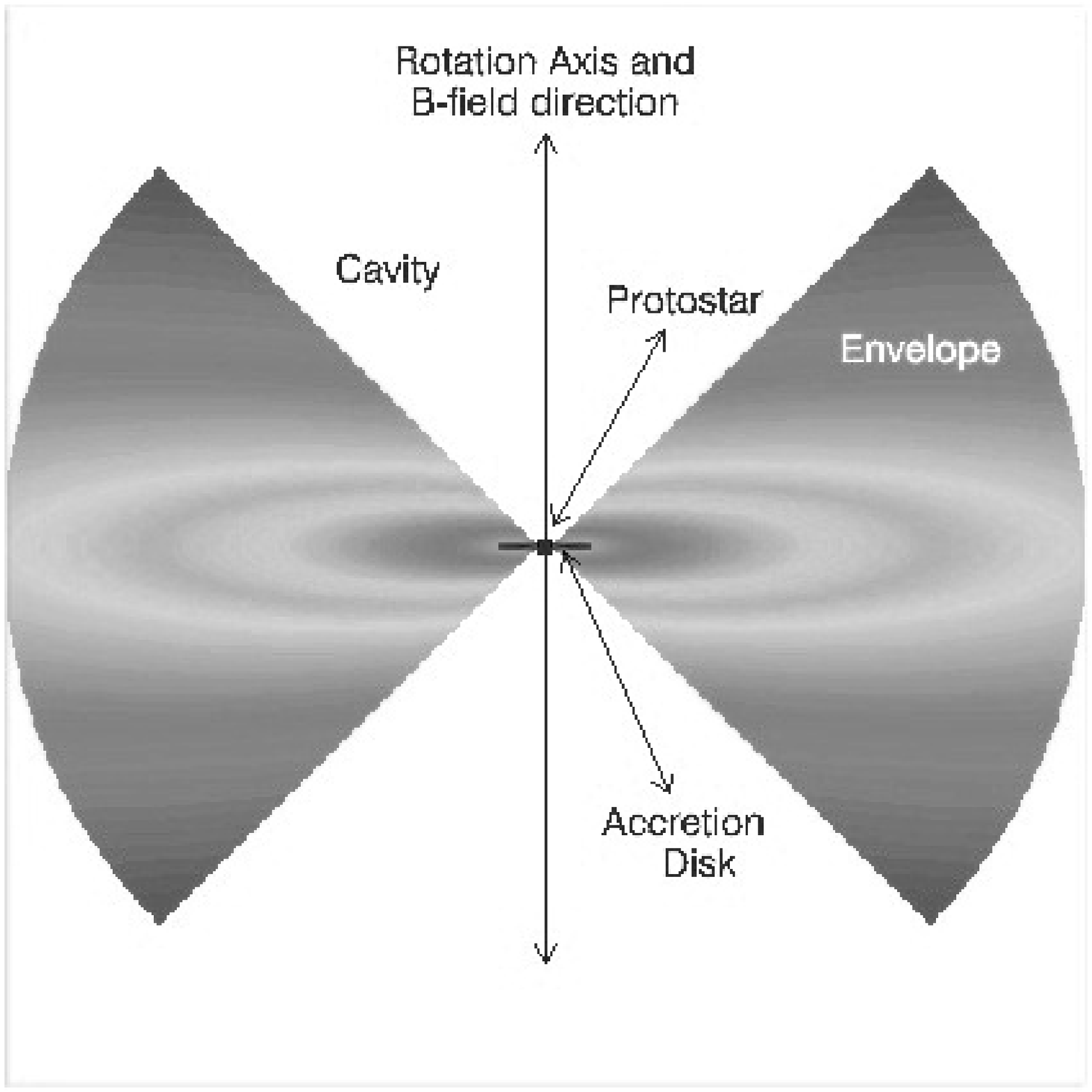}}

\put(-50,-69){\framebox(278,278)}

\end{picture} 
\end{center}
\vspace{3cm}
Figure 1
\end{figure*}

\pagebreak

\begin{figure*}[thbp]
\begin{center}
\begin{picture}(200,210)

\put(0,0){\includegraphics{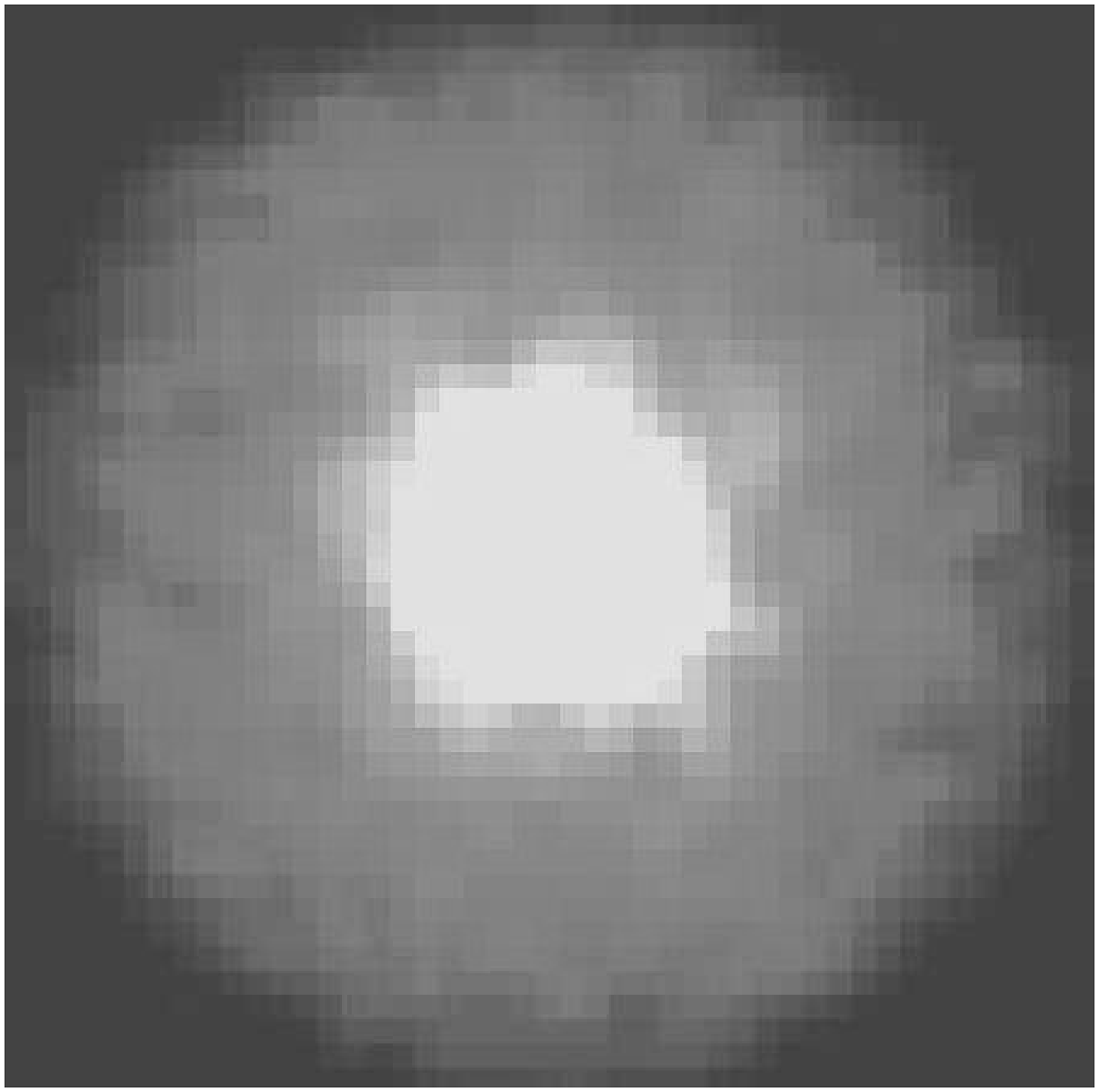}}

\put(0,0){\includegraphics{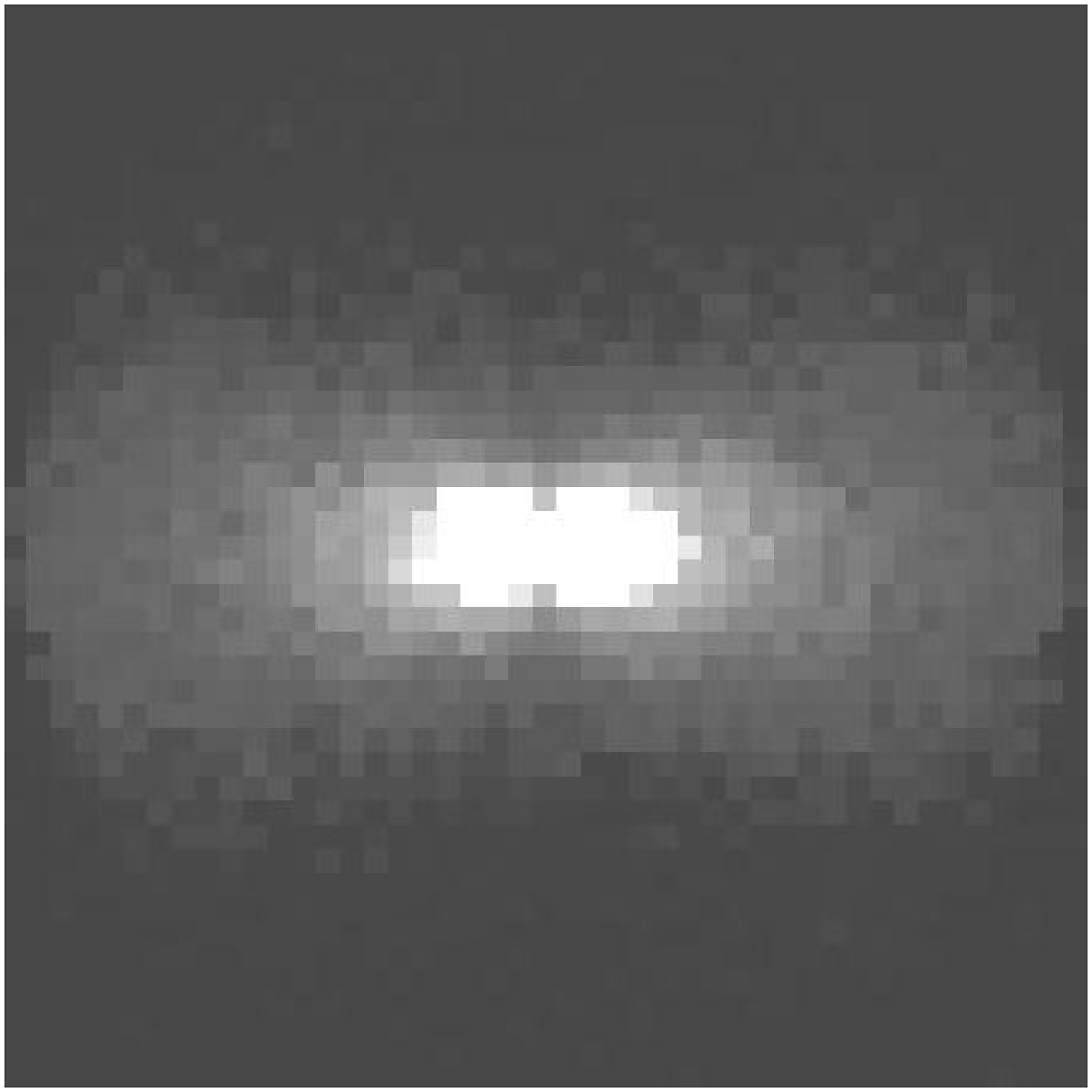}}

\put(-140,170){(a) Stokes I:  $\tau=0.5$}
\put(100,170){(b) Stokes I: $\tau=10$}
\end{picture} 
\end{center}
\vspace{3cm}
Figure 2(a-b)
\end{figure*}

\pagebreak
\begin{figure*}[thbp]
\begin{center}
\begin{picture}(200,210)

\put(0,0){\includegraphics{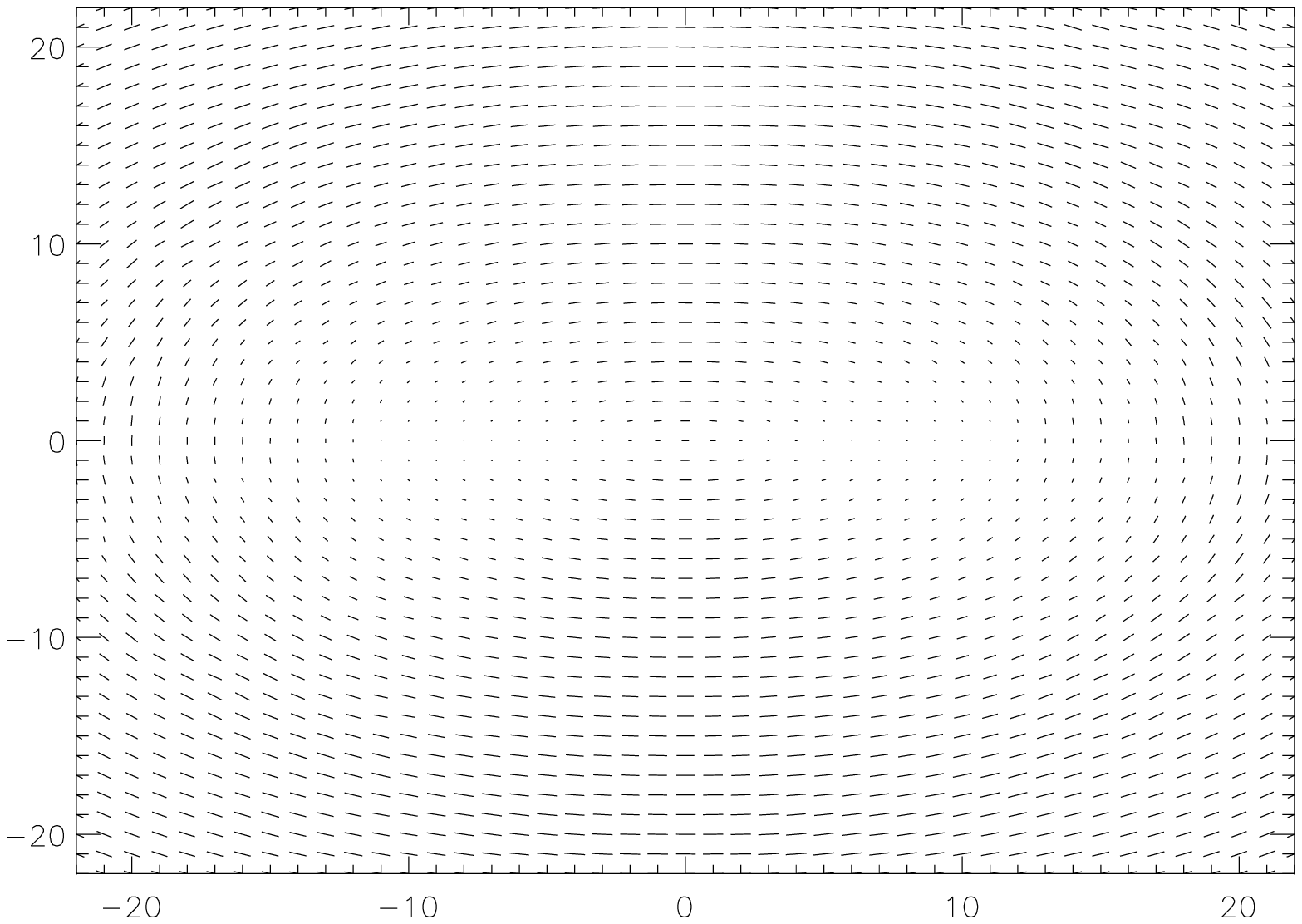}}

\put(0,0){\includegraphics{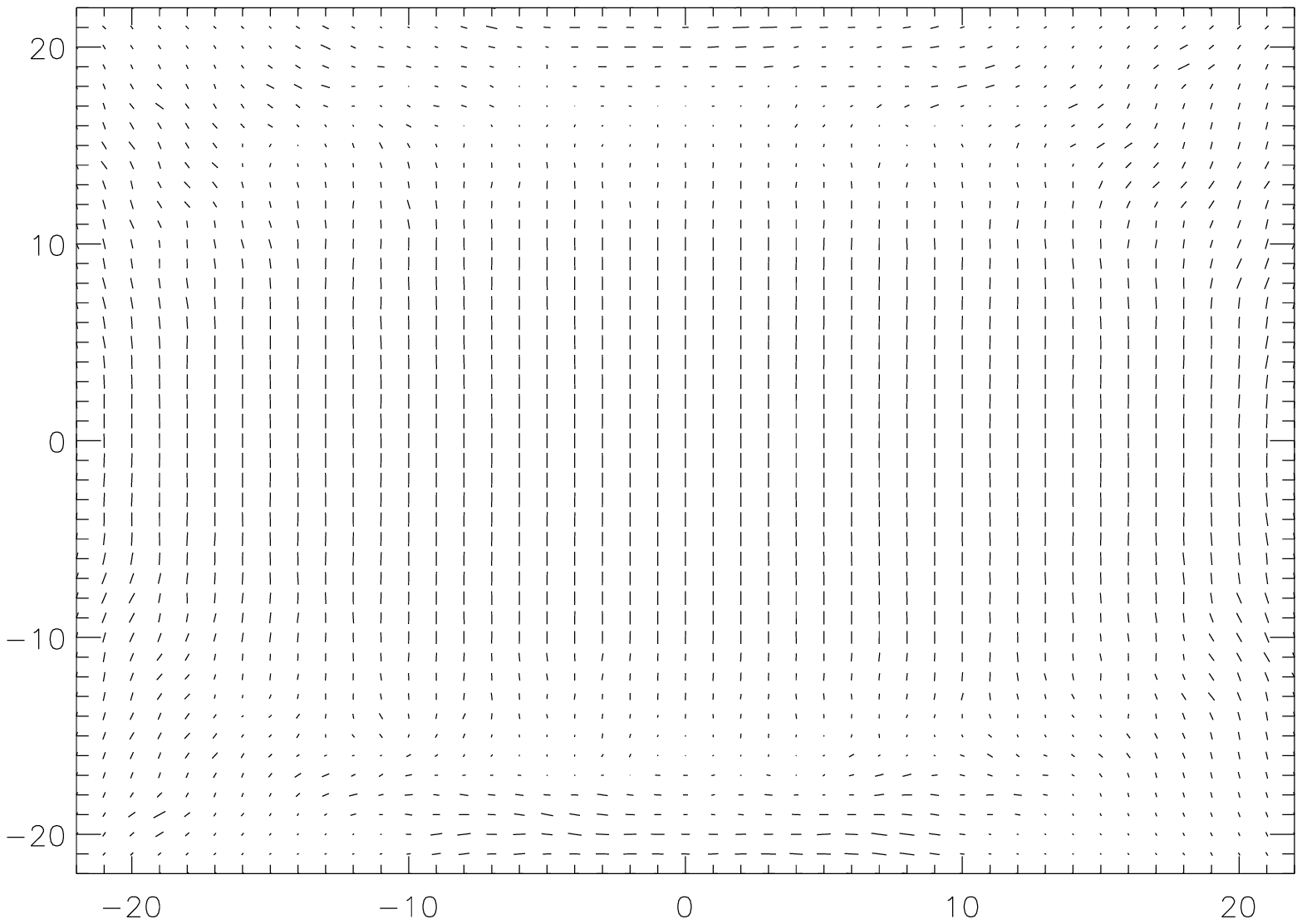}}

\put(-140,250){(a) $\tau=0.5$}
\put(120,250){(b) $\tau=5$}
\end{picture} 
\end{center}
Figure 3(a-b)
\end{figure*}

\pagebreak
\begin{figure*}[thbp]
\begin{center}
\begin{picture}(200,480)

\put(0,0){\includegraphics{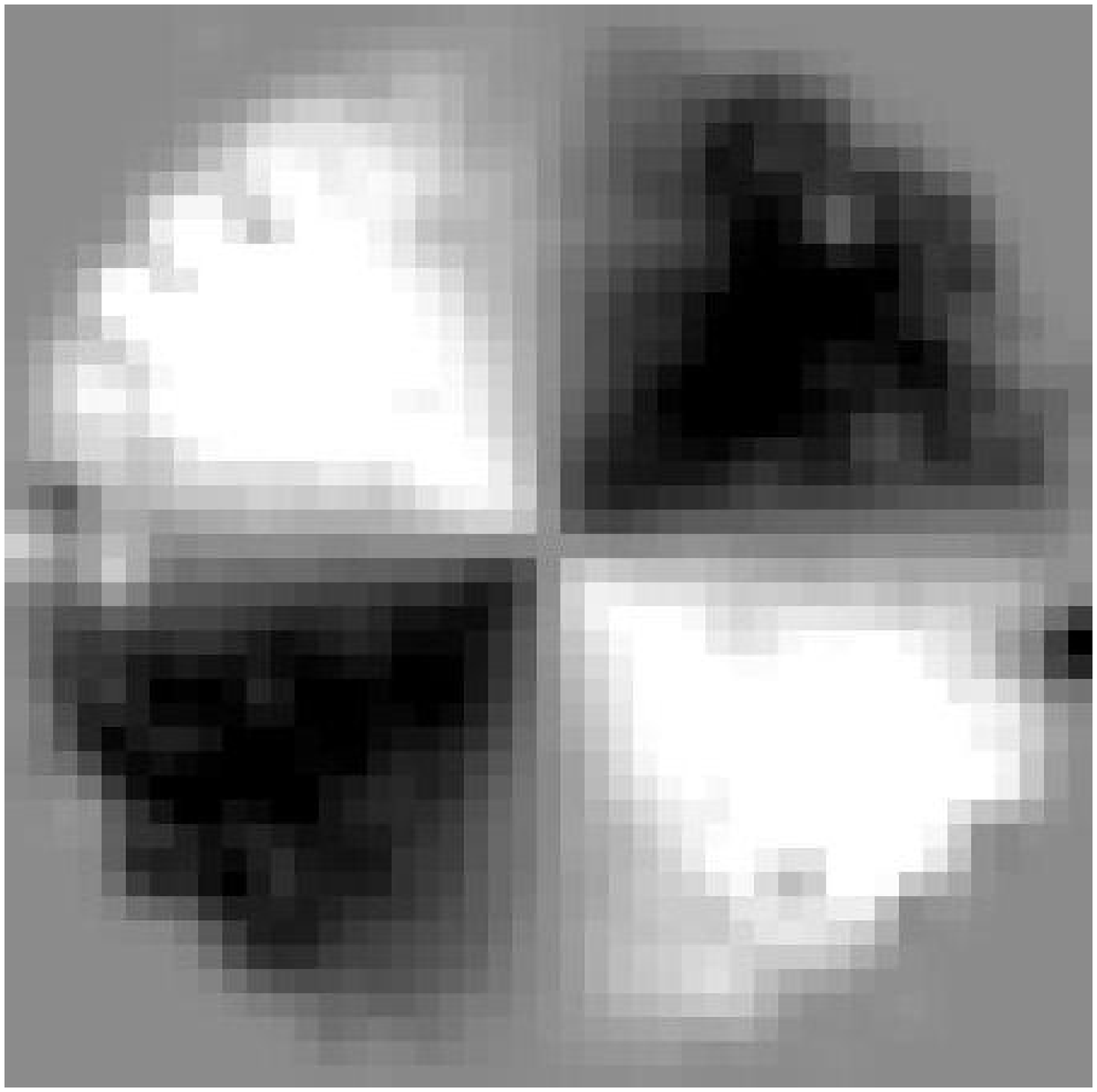}}

\put(0,0){\includegraphics{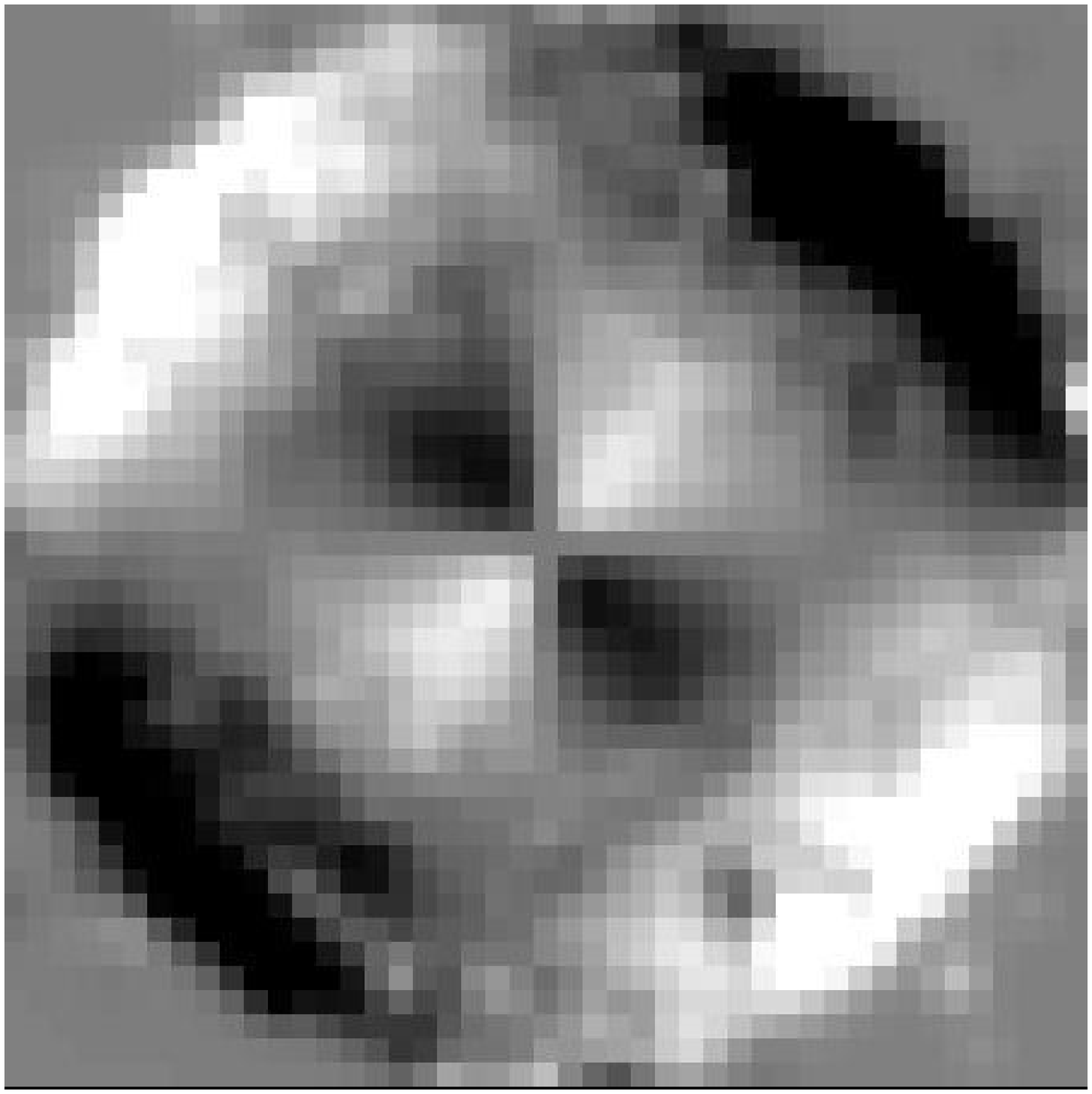}}

\put(0,0){\includegraphics{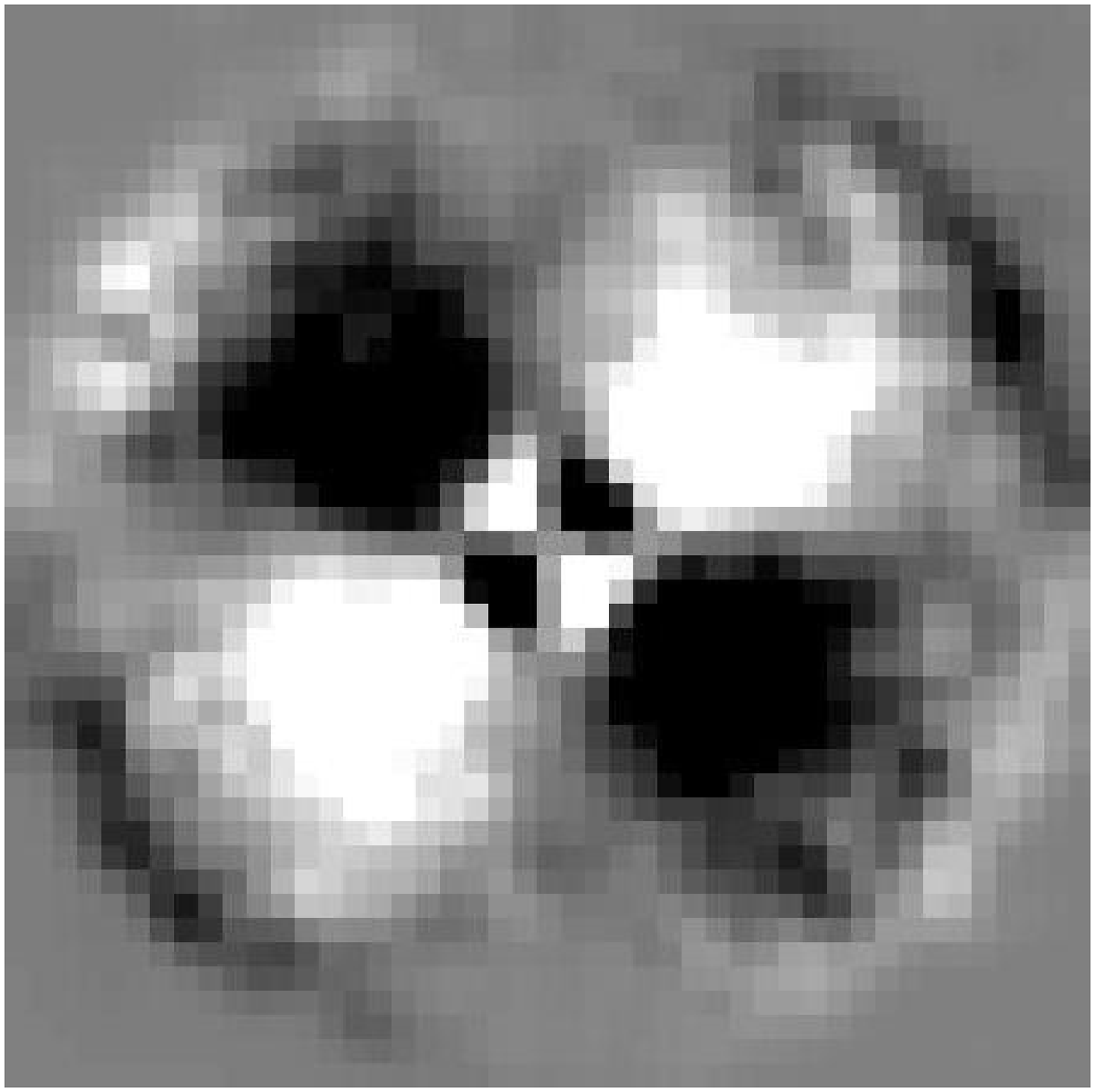}}

\put(20,540){{\bf V/I: $\tau=0.5$}}
\put(20,290){{\bf V/I: $\tau=3$}}
\put(20,35){{\bf U: $\tau=3$}}

\put(-60,540){(a)}
\put(-60,290){(b)}
\put(-60,35){(c)}

\end{picture} 
\end{center}
Figure 4(a-c)
\end{figure*}

\begin{figure*}[thbp]
\begin{center}
\begin{picture}(200,480)

\put(0,0){\includegraphics{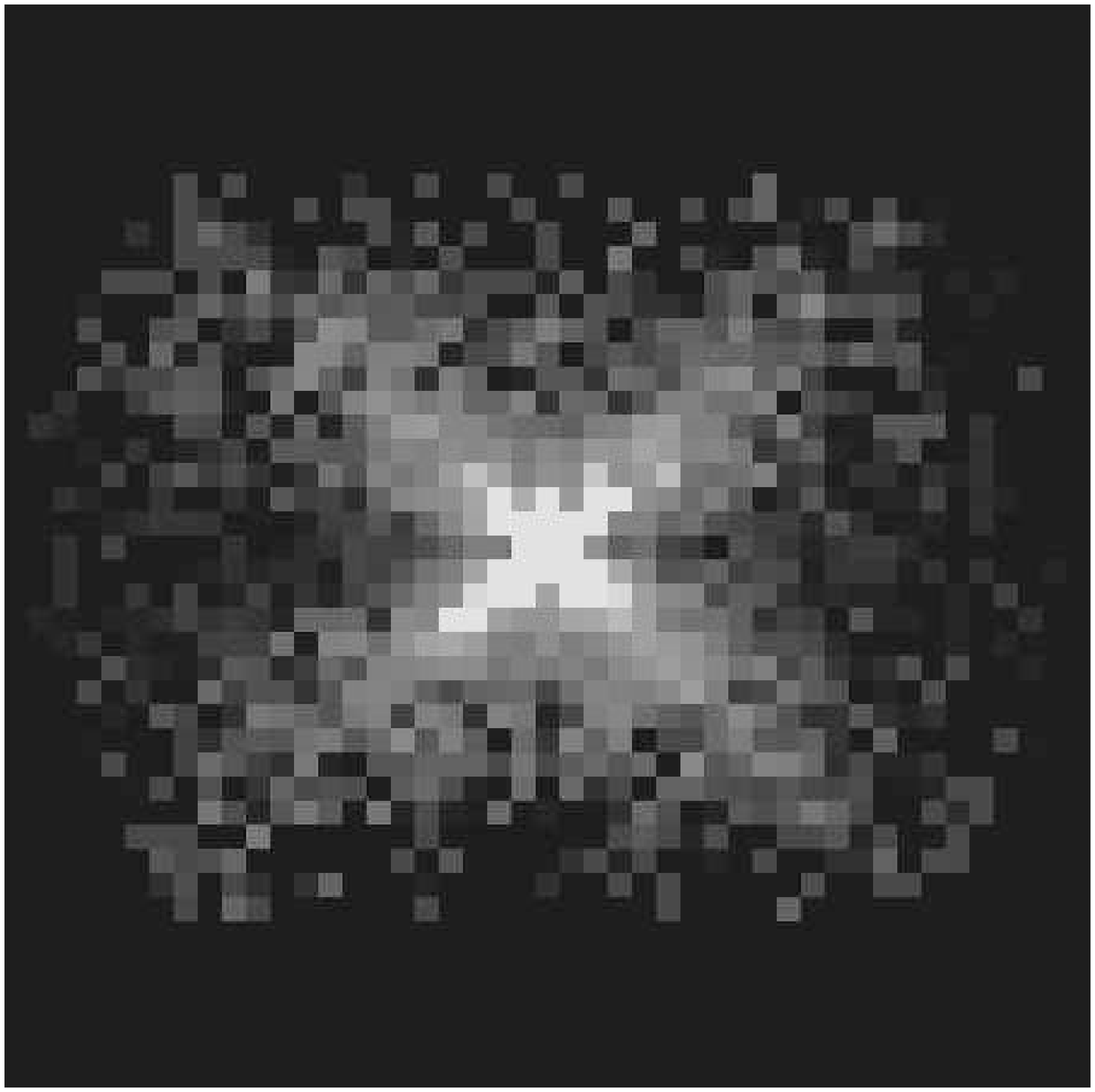}}

\put(0,0){\includegraphics{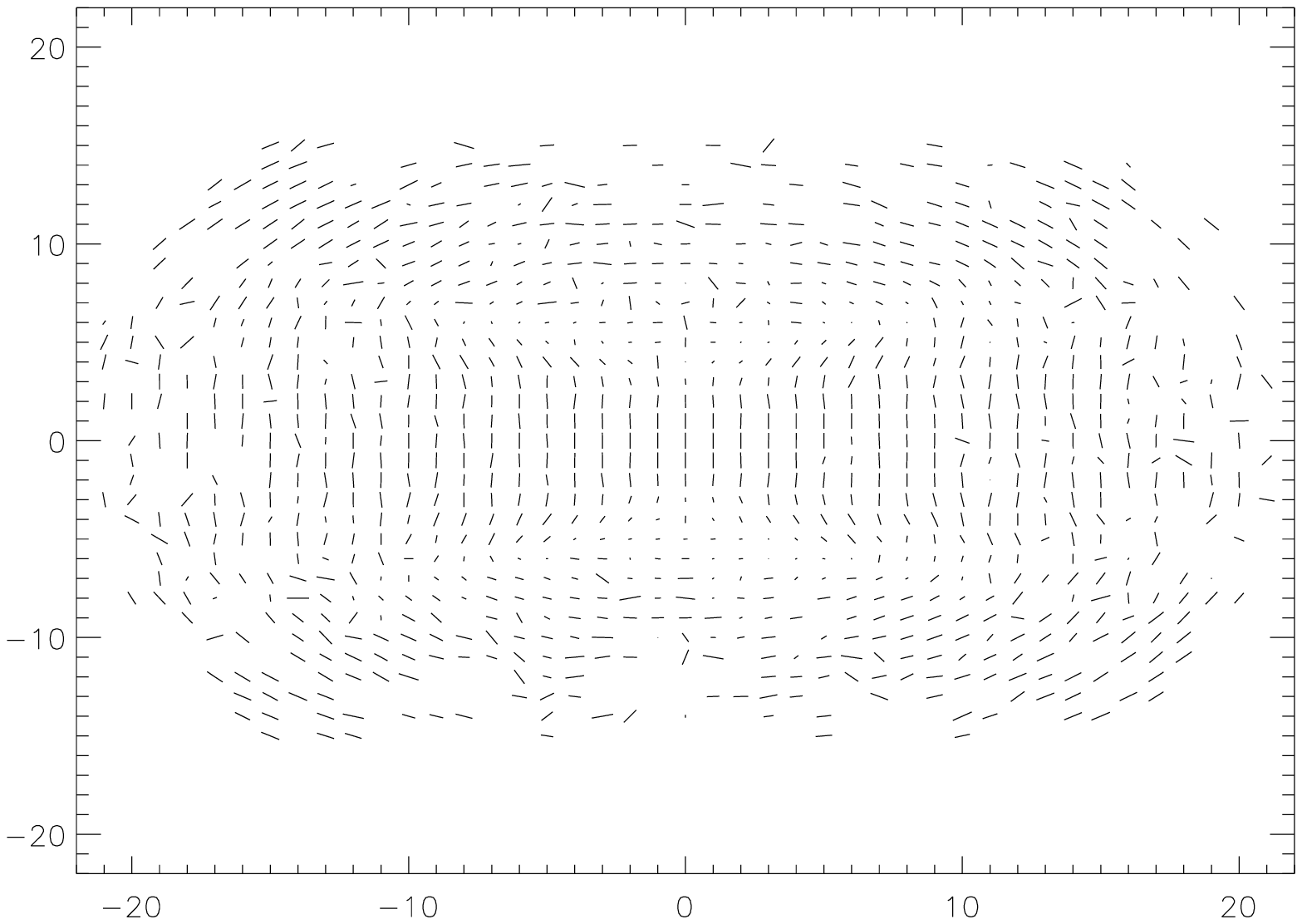}}

\put(0,0){\includegraphics{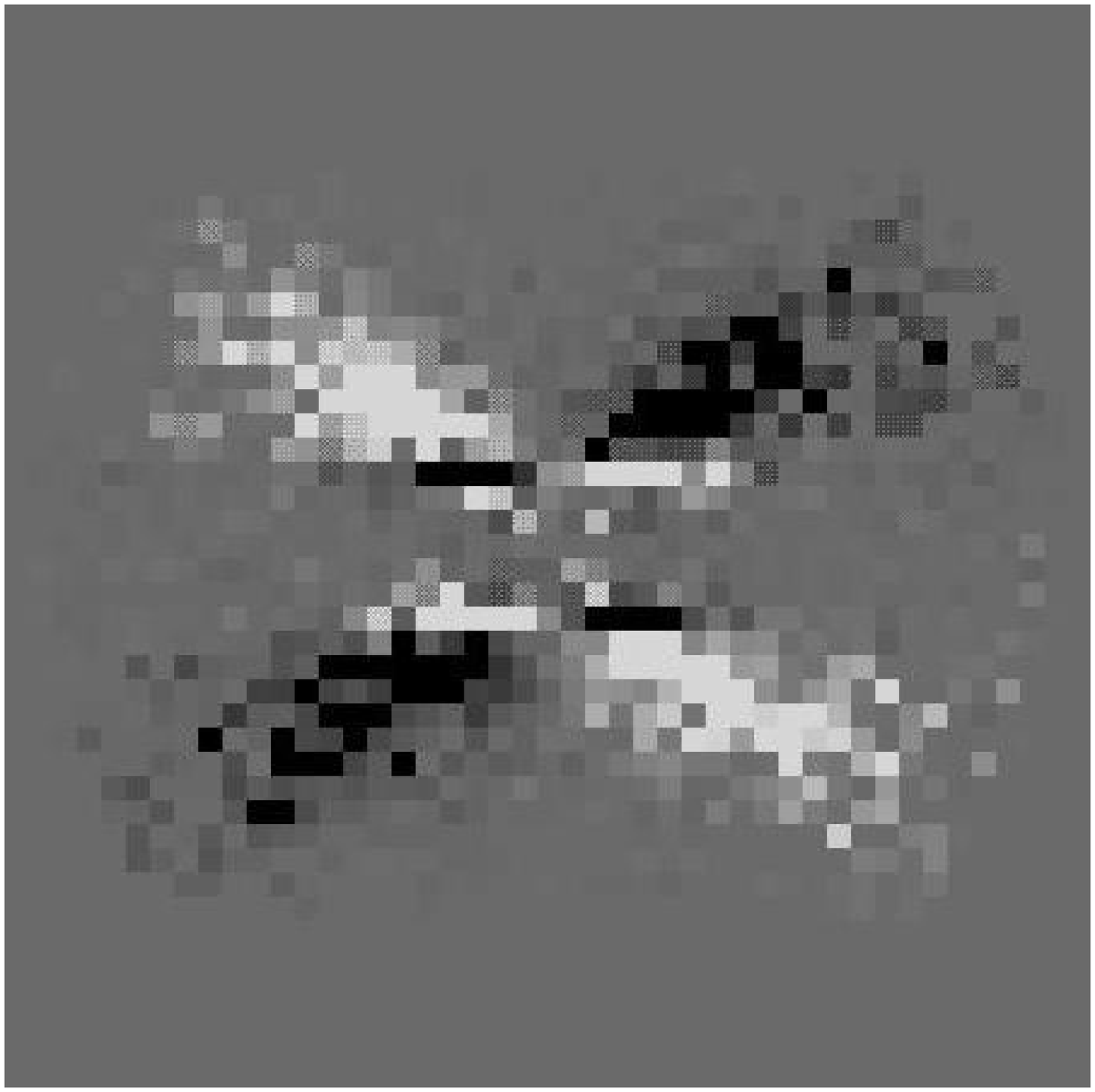}}

\put(0,0){\includegraphics{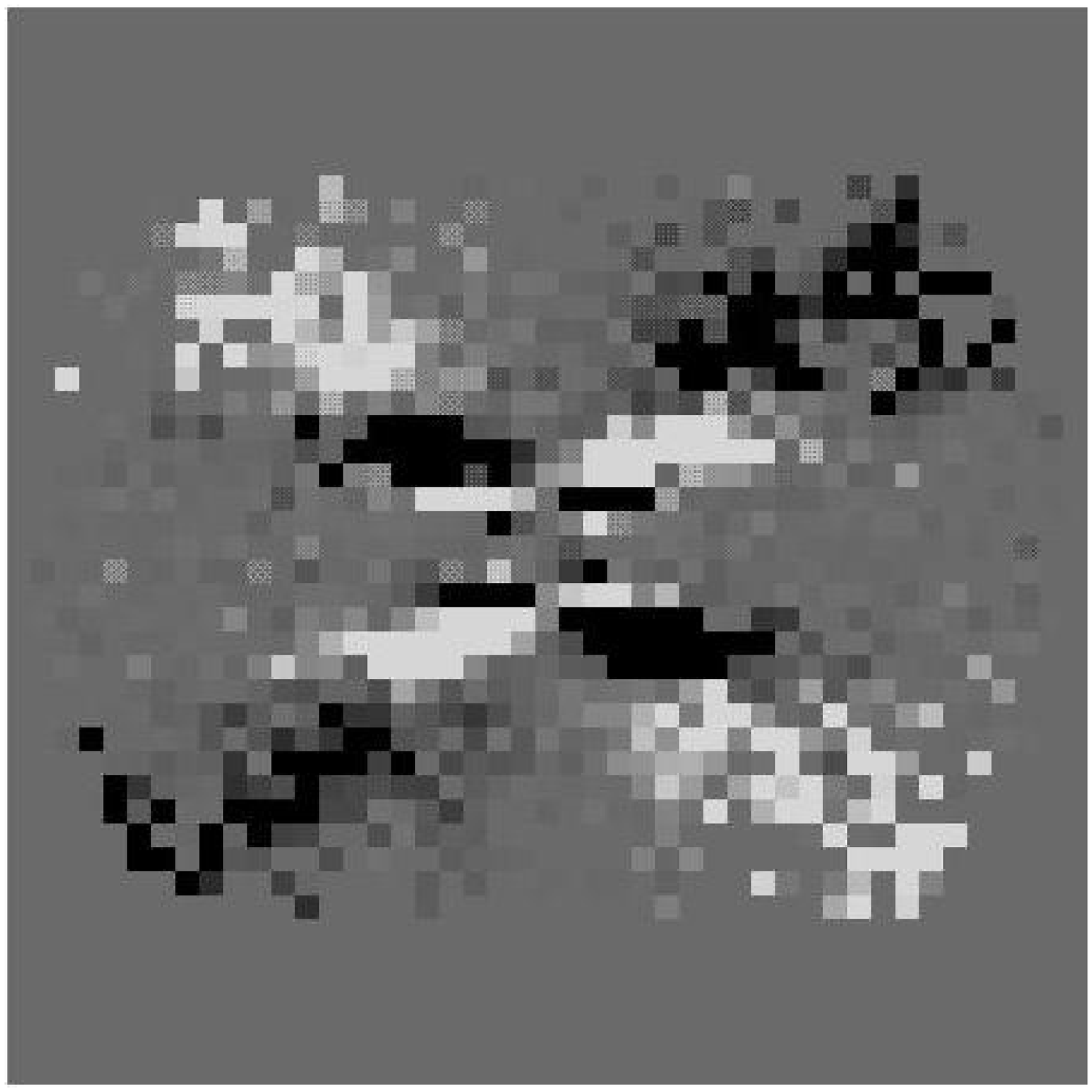}}

\put(-50,620){{\bf I}}
\put(-50,345){{\bf V}}
\put(200,345){{\bf U}}

\put(-150,615){(a)}
\put(100,615){(b)}
\put(-150,345){(c)}
\put(100,345){(d)}

\end{picture} 
\end{center}
Figure 5(a-d)
\end{figure*}

\begin{figure*}[thbp]
\begin{center}
\begin{picture}(200,400)

\put(0,0){\includegraphics{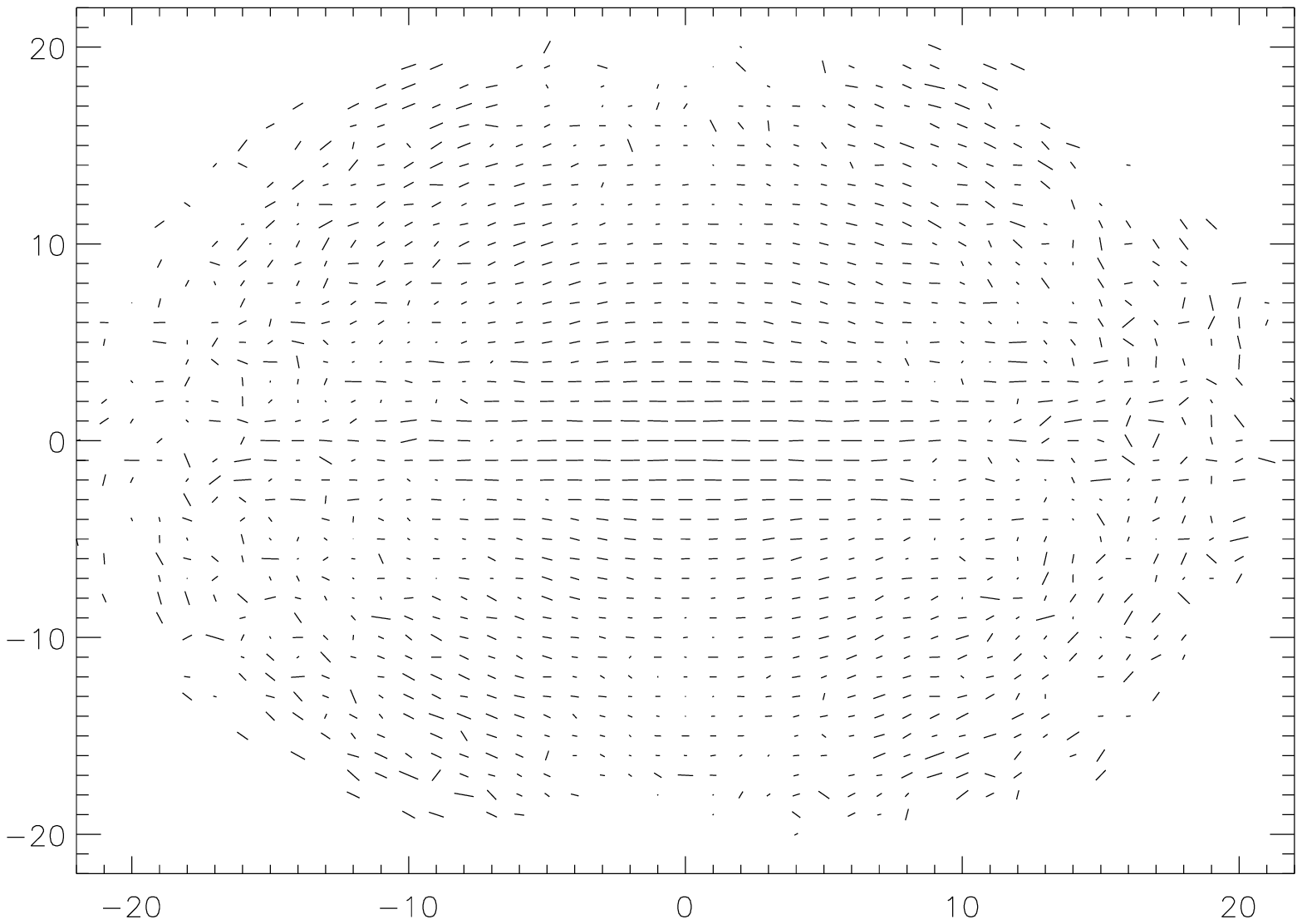}}

\end{picture} 
\end{center}
Figure 6
\end{figure*}

\pagebreak
\begin{figure*}[thbp]
\begin{center}
\begin{picture}(200,400)

\put(0,0){\includegraphics{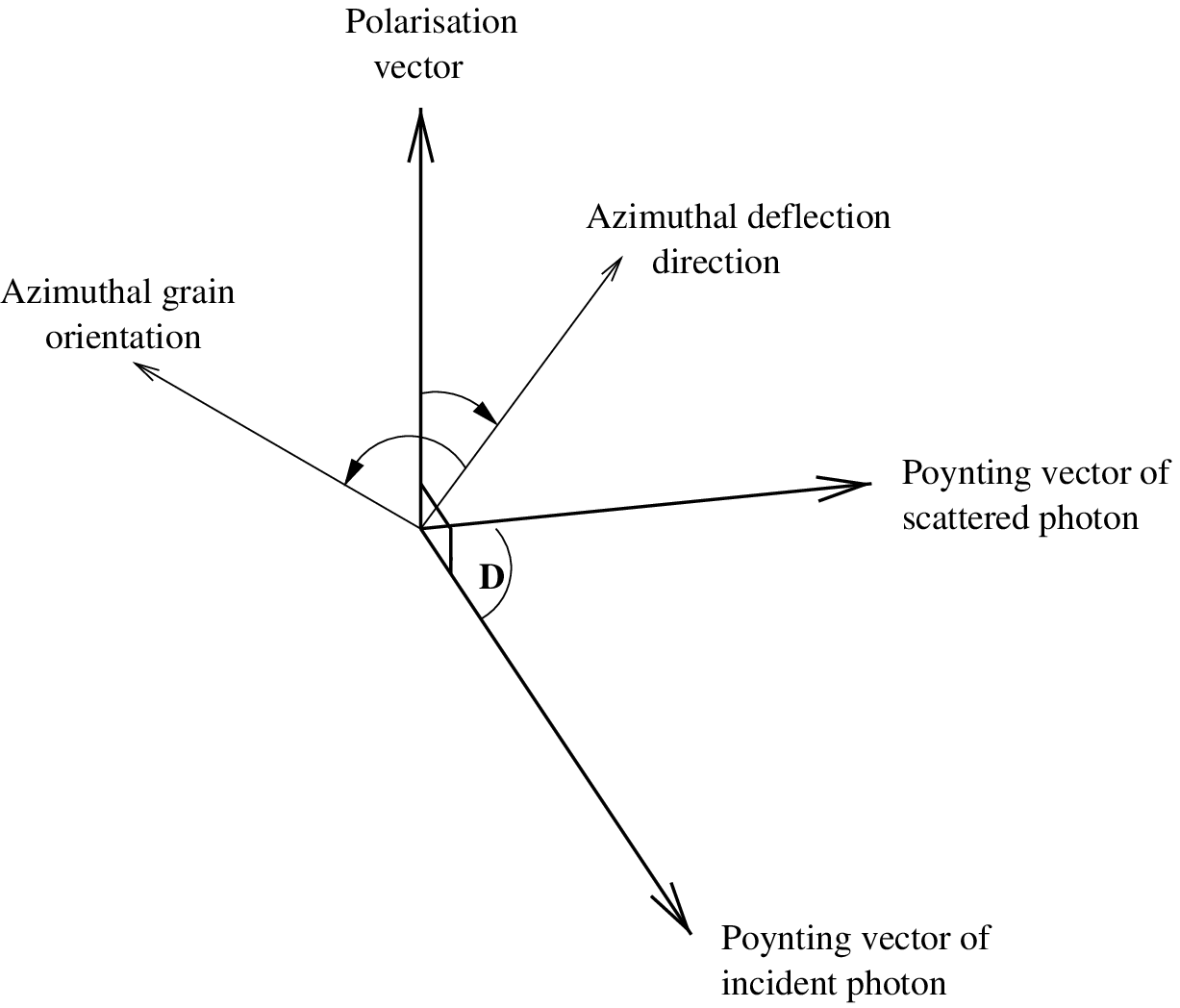}}

\put(25,140){$\gamma$}
\put(10,125){$\alpha$}

\end{picture} 
\end{center}
\vspace{2cm}
Figure 7
\end{figure*}

\end{document}